\documentclass[showpacs,preprintnumbers,amsmath,amssymb,superscriptaddress,nofootinbib]{revtex4}

\usepackage{axodraw4j}
\usepackage{pstricks}
\usepackage{color}
\usepackage{graphicx}
\usepackage{dcolumn}
\usepackage{bm}
\usepackage{atlasphysics}

\begin{document}

\title{
Higgs Boson Searches and the $Hb\bar b$ Coupling at the LHeC}
\author{Tao Han}
\email{than@hep.wisc.edu}
\author{Bruce Mellado}
\email{bmellado@wisc.edu}
\affiliation{Department of Physics, University of Wisconsin, Madison, WI 53706, USA}


\begin{abstract}
\vspace*{0.1cm}
Once the existence of the Higgs boson is established at the CERN Large Hadron Collider (LHC), 
the focus will be shifted toward understanding its couplings to other particles. A crucial aspect is the measurement of the bottom Yukawa coupling, which is challenging at the LHC.
In this paper we study the use of forward jet tagging as a means to secure the observation 
and to significantly improve the purity of the Higgs boson signal in the $H\rightarrow b\overline{b}$ decay mode from 
deep inelastic electron-proton scattering at the LHC. 
We demonstrate that the requirement of forward jet tagging in charged current events strongly enhances the signal-to-background ratio. The impact of a veto on additional partons is also discussed. Excellent response to hadronic shower and $b$-tagging capabilities are pivotal detector performance aspects. 

\end{abstract}


\maketitle

\section{Introduction}
\label{sec:intro}

In the Standard Model (SM) of electroweak interactions, 
%
the Higgs field is responsible for generating masses to all of the particles in the theory, the gauge bosons as well as
the fermions.
%
The observation of the Higgs boson is a priority at the Large Hadron Collider (LHC)~\cite{Evans:2008zzb} 
for the CMS~\cite{CMSPTDR,:2008zzk} and ATLAS~\cite{:2008zzm,Aad:2009wy} experiments in order to
test the mechanism for the electroweak gauge symmetry breaking.
Once the existence of the Higgs bosons is established, the focus will be shifted toward understanding its couplings to other particles, in particular to the fermions. While it seems to be feasible to observe the production $gg\to H$ 
to indirectly test the $Ht\bar t$ coupling, and the decay $H\to \tau\bar \tau$~\cite{Rainwater:1998kj}, the
Yukawa couplings to other fermions remain very difficult to access at the LHC. A 
crucial aspect is the measurement of the bottom Yukawa coupling~\cite{Zeppenfeld:2000td,Duhrssen:2004cv}
since it is pivotal to check the consistency of the SM and beyond. 
Extensive studies have been performed over the years to assess the feasibility of this measurement~\cite{RichterWas:1999sa,Drollinger:2001ym,Maltoni:2001hu,Drollinger:2002uj,Mangano:2002wn,Gabrielli:2007wf,Butterworth:2008iy}. Nevertheless, the observation of the $H\to bb$ decay remains very challenging at the LHC~\cite{CMSPTDR,Aad:2009wy}.

Recently, there has been a consideration for 
high energy $ep$ collisions with the Deep Inelastic Electron-Nucleon Scattering at the LHC 
(LHeC)~\cite{Dainton:2006wd}. The energy of the incoming proton is given by the LHC beam, 
and several scenarios are considered for the energy of the incoming electron as
\begin{equation}
E_p =7\ {\tev},\quad E_e=50-200\ {\gev}, 
\end{equation}
corresponding to the center of mass energies of $\sqrt s = 2\sqrt{ E_p E_e} \approx 1.18-2.37\,\tev$. 
The anticipated integrated luminosity is about the order of 10$-$100 fb$^{-1}$ 
depending on the energy of the incoming electron and the design. 

There will be a rich physics program at the LHeC \cite{UtaKleinDIS09}.
Several studies have been reported recently regarding the feasibility of the observation of a SM Higgs boson with the decay mode $H\rightarrow b\overline{b}$~\cite{UtaKleinDIS09,MasakiDIS09}. The prospects for observing this channel at the LHeC are very exciting.  By combining this measurement with the observation of $H\rightarrow WW^{\star},\tau\tau$ channels 
from the LHC we can expect to directly extract the bottom Yukawa coupling.
In this paper we explore this aspect in detail for the LHeC. We use the forward jet-tagging as a means to secure 
the feasibility of the observation by significantly improving the purity of the Higgs boson signal.
The implications of a central jet veto are also discussed. 
The Higgs boson signal efficiency is studied as a function of the energy of the incoming electron. 
The impact of the most relevant aspects of the detector performance is evaluated.

In Section~\ref{sec:higgsprod} we briefly summarize the characteristics of the Higgs boson production 
in $ep$ collisions for both the charged current (CC) and neutral current (NC) processes.
In Section~\ref{sec:background} we give a description of the background processes considered here. In Section~\ref{sec:kinematics} we quantify signal and background yields after the requirement of forward parton tagging. Results are summarized in Section~\ref{sec:conclusions}.

\section{Higgs Boson Production in High Energy $ep$ Collisions}
\label{sec:higgsprod}
 
\begin{figure}[t]
%
\begin{center}
\fcolorbox{white}{white}{
  \begin{picture}(213,176) (206,-27)
    \SetWidth{1.5}
    \SetColor{Black}
    \Line[arrow,arrowpos=0.5,arrowlength=8.333,arrowwidth=3.333,arrowinset=0.2](208,80)(304,96)
    \Line[arrow,arrowpos=0.5,arrowlength=8.333,arrowwidth=3.333,arrowinset=0.2](304,96)(400,128)
    \Line[arrow,arrowpos=0.5,arrowlength=8.333,arrowwidth=3.333,arrowinset=0.2](208,-16)(304,0)
    \Line[arrow,arrowpos=0.5,arrowlength=8.333,arrowwidth=3.333,arrowinset=0.2](304,0)(416,0)
    \Photon(304,0)(352,48){7.5}{3}
    \Photon(304,96)(352,48){7.5}{3}
    \Line[dash,dashsize=10](352,48)(416,64)
    \Text(336,128)[lb]{\Large{\Black{$\nu_e\ (e^\pm)$}}}
    \Text(224,96)[lb]{\Large{\Black{$e^{\pm}$}}}
    \Text(271,60)[lb]{\Large{\Black{$W^\pm (Z)$}}}
    \Text(271,16)[lb]{\Large{\Black{$W^\mp (Z)$}}}
    \Text(256,-32)[lb]{\Large{\Black{$q$}}}
    \Text(368,-32)[lb]{\Large{\Black{$q^{\prime}\ (q)$}}}
    \Text(384,80)[lb]{\Large{\Black{$H$}}}
  \end{picture}
}
\end{center}
\caption{Leading order diagram for the production of a Standard Model Higgs boson  in $ep$ collisions 
for the charged current and neutral current processes.
  \label{fig:vbfh}}
\end{figure}
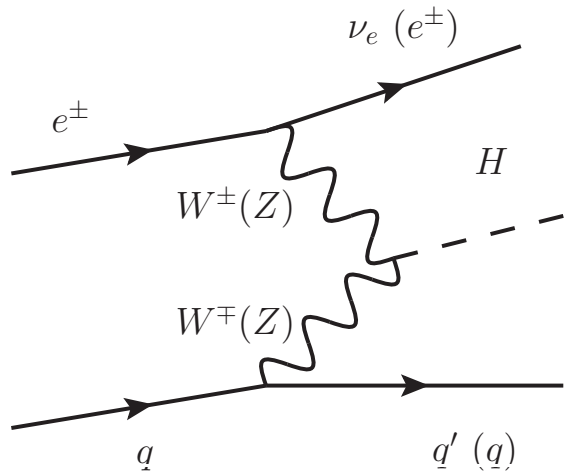

The leading production mechanism for the SM Higgs boson at the LHeC is 
\begin{equation}
eq\rightarrow \nu_e Hq^{\prime} \quad {\rm and}\quad
eq\rightarrow e Hq, 
\label{eq:process}
\end{equation}
via Vector Boson Fusion processes (VBF), as depicted in Fig.~\ref{fig:vbfh}. 
It is remarkable that the Higgs boson production via VBF was first calculated for lepton-nucleon interactions
\cite{Ellis:1975ap,LoSecco:1976ii,Godbole:1977ti,
Hioki:1983yz,Han:1985zn}. Kinematic features of the signal were exploreed in~\cite{Grindhammer:1990vs}.

 %

\begin{figure}[tb]
\vspace*{0.2cm}
\centerline{\includegraphics[height=6.5cm,angle=0]{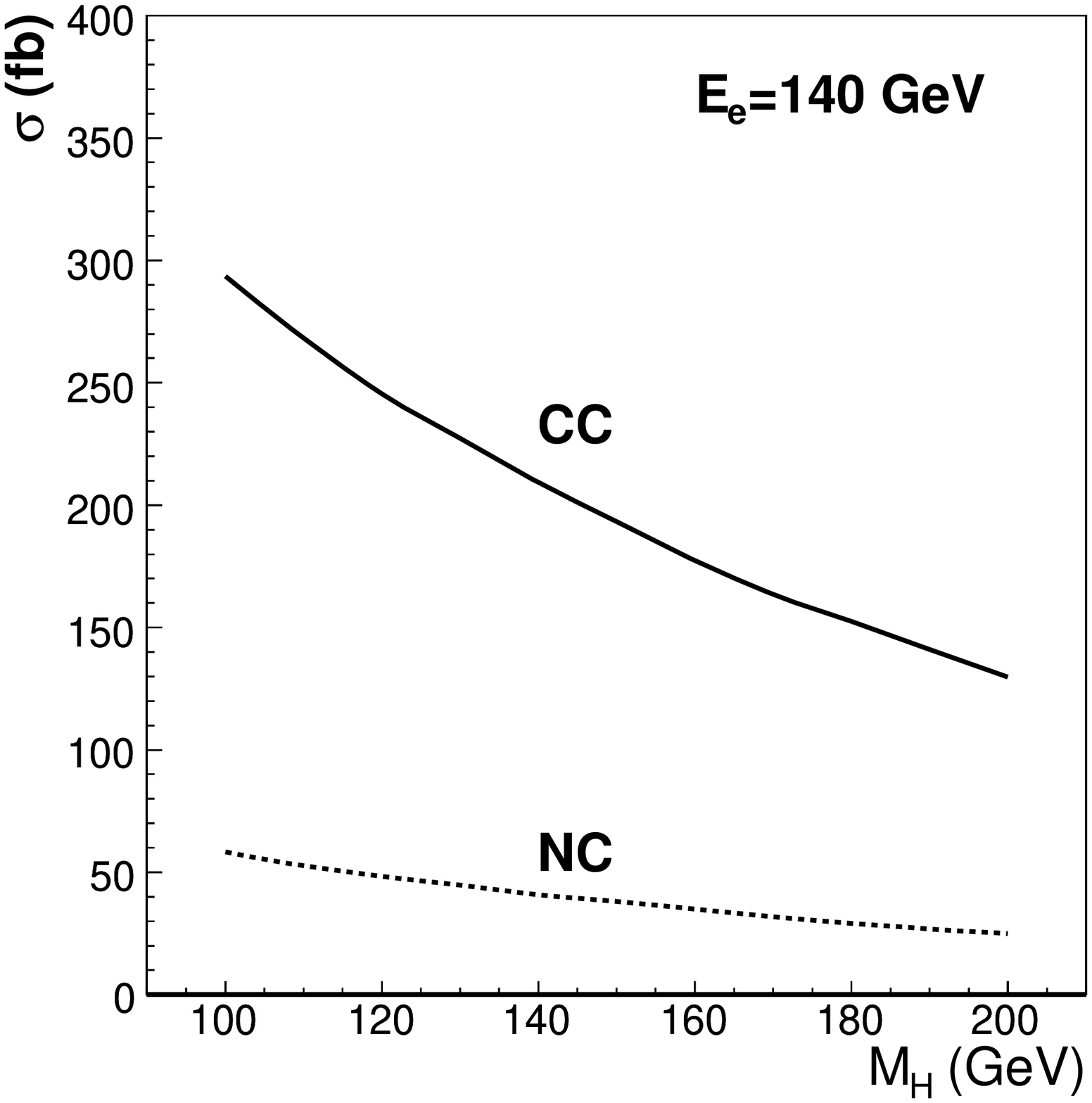}
\includegraphics[height=6.5cm,angle=0]{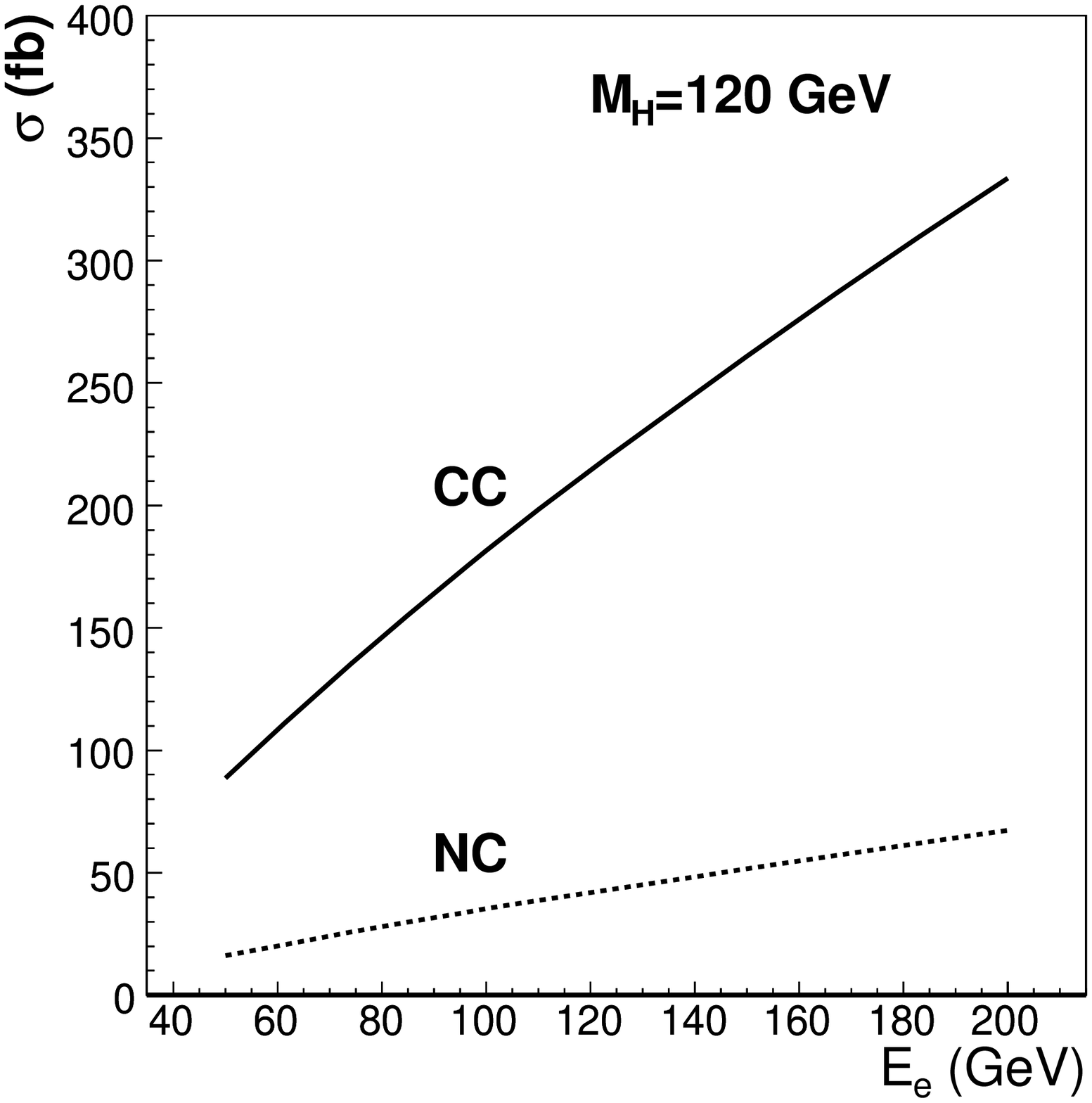}
}
\caption{Total cross sections, in fb, for the SM Higgs boson production in $ep$ collisions at the for both CC (solid curves) and NC (dashed curves) processes. The plot on the left (a) shows the mass dependence for a fixed value of the energy of the 
incoming electron, $E_e=140\,\gev$. The plot on the right (b) dispays the $E_e$ dependence for a fixed Higgs  boson mass, 
$M^{}_H=120\,\gev$.
  \label{fig:tot}}
\end{figure}

We first present the total cross sections  in $ep$ collisions  as a function of the c.m.~energy
and the Higgs boson mass $M_H$ without any acceptance cuts. 
Figure~\ref{fig:tot} displays the dependence of the cross-section (a) on the Higgs boson mass and 
(b) on the energy 
of the inconing electron. We see that the production rate for the CC process (solid curves) 
is larger than that of the NC process (dotted curves) by about a factor of $4-6$. 
This is mainly due to the accidentally suppressed NC coupling to the electrons. 
Here we have used the package MadGraph \cite{Alwall:2007st}
for the full matrix element calculations at tree-level, 
and adopted the parton distribution functions CTEQ6L1~\cite{Pumplin:2002vw}. 
We choose the renormalization and factorization scales to be at the $W$-mass, which characterizes the typical
momentum transfer for the signal processes. 

\begin{figure}[tb]
\vspace*{0.2cm}
\centerline{
\includegraphics[height=6.5cm,angle=0]{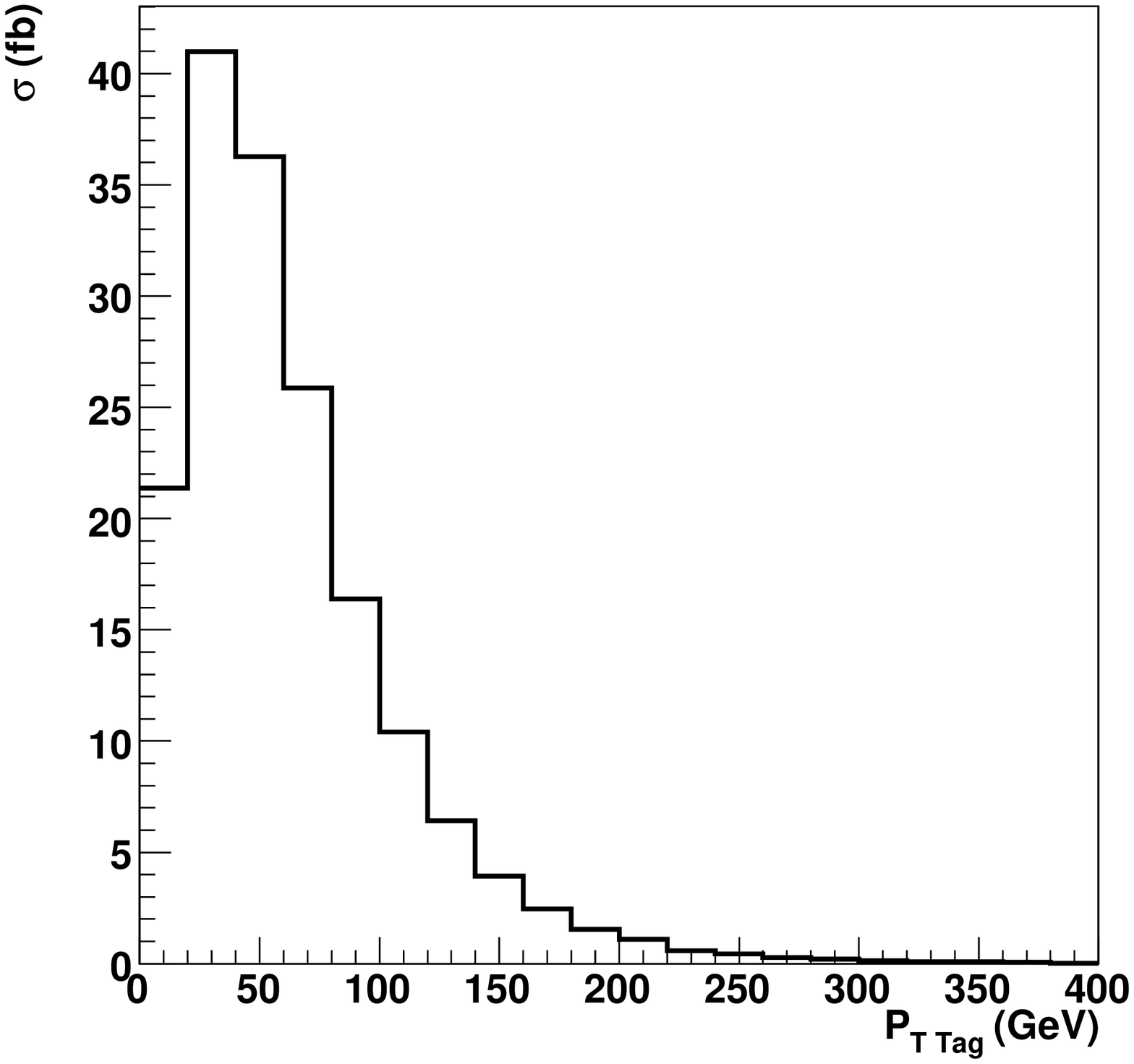}
\includegraphics[height=6.5cm,angle=0]{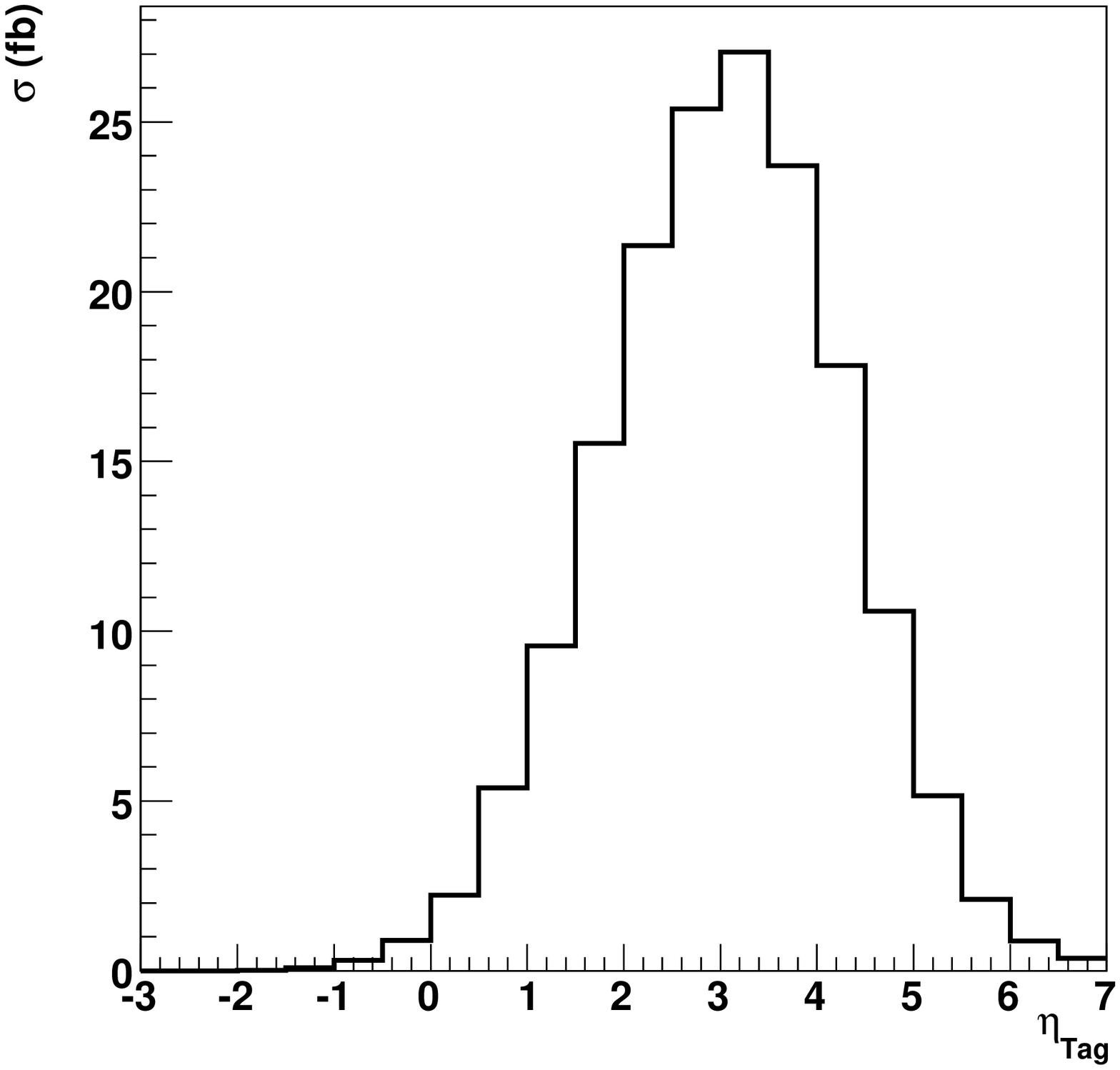} }
\caption{Kinematical distributions for the forward parton ($j$) in units of fb per bin
for the process  $e^-p\to \nu_e H j$.
The plot on the left (a) is for the transverse momentum distribution, and that on the right (b)
shows the  pseudo-rapidity distribution. 
  \label{fig:dist} }
\end{figure}

In order to appreciate the unique kinematics of the VBF process it is most intuitive to express the cross section
in a factorized form. Consider a fermion $f$ of a c.m.~energy $E$ radiating a gauge boson $V$ ($s \gg M_V^2$), the 
cross section of the scattering $fa\rightarrow f^{\prime} X$ via $V$ exchange can be expressed as:
\begin{equation}
\sigma(fa\rightarrow f^{\prime} X) \approx \int dx\ dp_T^2\ P_{V/f}(x,p_T^2)\ \sigma(Va\rightarrow X)
\label{eq:effWapp}
\end{equation}
where $\sigma(Va\rightarrow X)$ is the cross-section of the $Va\rightarrow X$ scattering and $P_{V/f}$ 
can be viewd as the probability distribution for a weak boson $V$ of energy $xE$ and transverse momentum $p_T$.
The dominant kinematical feature is a nearly collinear radiation of $V$ off $f$, often called 
the ``Effective $W$-Approximation" \cite{Cahn:1983ip,Chanowitz:1984ne,Kane:1984bb}.
when the center of mass energy is much greater than the mass of the weak bosons, the
probability distributions of the weak bosons with different polarizations can be approximated by
\begin{eqnarray}
P_{V/f}^T(x,p_T^2) &=&
{g_V^2 + g_V^2 \over 8\pi^2} {1 + (1-x)^2 \over x } { p_T^2 \over \left(p_T^2 + (1-x)M_V^2\right)^2 }
\label{eq:PT} \\
P_{V/f}^L(x,p_T^2) &=&
{g_V^2 + g_V^2 \over 4\pi^2} {1-x \over x } { (1-x) M_V^2 \over \left(p_T^2 + (1-x)M_V^2\right)^2 }.
\label{eq:PL}
\end{eqnarray}
%
These expressions lead us to the following observations:
\begin{itemize}
\item[1]
Unlike the QCD partons that scale like $1/p_T^2$ at the low 
transverse momentum, the final state quark $f'$ typically has 
$p_T\sim\sqrt{1-x}M_V\leq M_W$.

\item[2]  Due to the $1/x$ behavior for the gauge boson distribution, the out-going 
parton energy $\left(1-x\right)E$ tends to be high. Consequently, it leads to an energetic  forward jet
with small, but finite, angle with respect to the beam.

\item[3] At high $p_T$, $P_{V/f}^T\sim 1/p_T^2$ and $P_{V/f}^L\sim 1/p_T^4$, and thus 
the contribution from the longitudinally polarized gauge bosons is relatively suppressed 
at high $p_T$ to that of the transversely polarized. 

\end{itemize}
These features are illustrated in Fig.~\ref{fig:dist},  where the plots on the left and right display the 
transverse momentum and  pseudo-rapidity distributions of the out-going jet ($j$), 
respectively, for the process of $e^-p\to \nu_e H j$. Here and henceforth, for illustration, 
we have taken the parameters as
\begin{equation}
E_p = 7\  {\tev},\quad E_e = 140\ {\gev},\quad M_H=120\ {\gev}.
\label{eq:para}
\end{equation}
Supported by these figures, 
items 1 and 3 clearly motivate a tagging for a forward  jet to separate the QCD backgrounds~\cite{Kleiss:1987cj,Barger:1988mr},
while item 3 suggests a veto of central jets with high $p_T$ to suppress the backgrounds initiated 
from the transversely polarized gauge bosons, 
and from other high $p_T$ sources such as top quarks~\cite{Barger:1990py}.

Large samples of events were generated with the Madgraph package without making the Effective $W$-Approximation. 
The branching fraction to $b\overline{b}$ is obtained with HDECAY~\cite{Djouadi:1997yw} 
and it is equal to 0.677 for $M_H=120$ GeV. 
The kinematics of the Higgs boson decays are handled by the decay interface of the Madgraph package.

We also tempt to consider the process of the double Higgs boson production $e^-p\rightarrow \nu_e HHj+X$.
With same settings as for the single Higgs boson production in Eq.~(\ref{eq:para}), the 
cross-section is about 0.05\,fb. 
With an increase of the energy of the incoming electron to 500\,\gev the cross-section would increase to 0.3\,fb.
The rather small cross section makes a signal observation difficult. In particular, it renders 
setting meaningful limits on the trilinear Higgs boson self-coupling extremely challenging. We will thus not consider
this double HIggs boson process further in this work.

\section{Charged Current Signal}

Because of the large production rate, we focus on the Higgs boson production signal via the charged current (CC)
process in  Eq.~(\ref{eq:process}). The signal topology consists of large missing transverse energy, a forward energetic jet, and $b\bar b$ from the Higgs decay.

\subsection{Background Processes}
\label{sec:background}

The two groups of leading backgrounds under consideration are the charged current background, 
and the photo-production.  The CC processes considered are generically expressed as
\begin{equation}
e^-p\rightarrow \nu_e\  q q'\ j +X.
\label{eq:cc}
\end{equation}
These include contributions from subprocesses $Wg \to t\bar b,\ qq';\   WW\to b\bar b;\ WZ/\gamma^*\to  qq'$ etc.
%
The photo-production processes considered are 
\begin{equation}
e^-p\rightarrow e^- \ q q' \ j+X.
\label{eq:nc}
\end{equation}
They include the subprocesses $\gamma g\to b \bar b;\ t\bar t$ etc. 
We do not consider the neutral current backgrounds with momentum transfer $Q^2>1\,\gev$.
Because of the presence of a final state electron at a large angle, 
it is assumed that these events can be rejected by tagging the scattered electron 
and/or requiring cuts on the $E-p_Z$ in the event. We did not include photo-production backgrounds involving resolved photons. We do not expect that this process renders a leading background.
Large missing transverse momentum thus arises in these processes primarily 
from mis-measurement of hadronic jets (see Section~\ref{sec:kinematics}) and 
the leptonic decays of the $W'$s.  

Cross-sections were calculated and events generated with the software package Madgraph~\cite{Alwall:2007st}. 
The top quark and $W$-boson decays were handled by the decay interface of the Madgraph package. 
The factorization and renormalization scales were set to the $Z$ boson mass.
We impose the following generator-level acceptance cuts
\begin{equation}
p_T(j) > 15\ \gev,\quad \Delta R >0.4.
\label{eq:gen}
\end{equation}
A summary of the cross-sections is given in the first row of Table~\ref{tab:evsel}. 

%
%

\subsection{Event Selection}
\label{sec:kinematics}

%
%
%

Several  experimental factors that contribute heavily to the feasibility of the Higgs boson search are considered here. First, in order to disentangle CC events from photo-production a good reconstruction of the missing transverse momentum is required. 
We evaluate the emergence of fake \met from mis-measurements of the energy of quark and gluons in the final state. This is performed by smearing the partonic energies with the hadronic energy relative resolution 
\begin{equation}
{\sigma_E \over E } = { \alpha \over \sqrt{E}} \oplus \beta, 
\label{eq:res}
\end{equation}
where we choose to take $\alpha=0.6\,\gev^{1 \over 2}$ and $\beta=0.03$. 
The resolution of the invariant mass of the Higgs boson candidate without smearing the angular reconstruction of the hadronic jets is about $7\%$. 
Checks have been performed with different values of $\alpha$ and $\beta$, see the next sub-section. 
The impact of hadronization and of the proton break-up on the 
$\met$ reconstruction are not taken into account here. 
Although we emphasize  the search for a Higgs boson in association with a forward parton, 
we adopt the same energy resolution as used above for other central jets for simplicity.

The ability to tag hadronic jets from $b$-quarks with high efficiency while displaying strong rejections against hadronic jets arising from lighter quarks is also a crucial experimental aspect. We assume a $b$-jet tagging efficiency of 
\begin{equation}
\epsilon_b = 60\,\%\quad {\rm in\ the\ range}\ \  \left|\eta\right|<2.5.
\end{equation}
The fake rejection factors of $10$ and $100$ are taken for c-jet and light jets, respectively. 

%

\subsection{CC Production}
\label{sec:cc}

The leading background to our Higgs signal are the charged current processes as in Eq.~(\ref{eq:cc}).
The following event selection is chosen:\footnote{The event selection shown here is not the result of an optimization procedure.}
\begin{itemize}
\item [\bf a] Require the presence of two $b$-partons with $p_T>30\,\gev$ in the pseudorapidity range $\left|\eta\right|<2.5$. 
The two $b$-partons constitute a Higgs boson candidate. To suppress photo-production backgrounds a cut on the missing transverse momentum, $\met>25\,\gev$ is required. The minimum difference in azimuthal angle between the observed $\met$ and the three partons in the event is required to be greater than 0.2\,rad. At this stage a charged lepton ($e, \mu, \tau$) veto with $p_T>10\,\gev$ in the range $\left|\eta\right|<2.5$ is applied.
\item [\bf b] The invariant mass of the two $b$-partons is required to be within 10\,\gev of the Higgs boson mass.
\item [\bf c] It is required that the leading remaining parton in the event have $p_T>30\,\gev$ and be found in the range 
$1<\eta<5$. This parton is referred to as the forward tagging parton.
\item [\bf d] The invariant mass of the Higgs boson candidate and the forward tagging parton, $M_{HJ}>250\,\gev$~\cite{Mellado:2004tj}.
\end{itemize}

\begin{table}[t]
\begin{center}
\begin{tabular}{|c||c||c|c|c||c|c||c||}
\cline{3-7}
\multicolumn{1}{c}{} & \multicolumn{1}{c|}{} & \multicolumn{3}{|c||}{CC} & \multicolumn{2}{c||}{Photo-prod.}  \\
\hline
Cuts & Higgs & $ t\overline{b}$ & $b\overline{b}j$ & $jjj$ & $b\overline{b}j$ & $t\overline{t}$ & $S/B $\\ 
\hline
Generator level  & 170  & 3800 &  810 &   26000 &   48000 &   250 &  -  \\
\hline \hline
{\bf a}  &  28 & 150 &  86 &   3.8 &   6.9 &   2.3 &   0.11
\\ 
{\bf b}  &  22 &  20 &   2.4 &   0.36 &   0.67 &   0.27 &   0.93
\\ 
{\bf c}  &  16 &   8.1 &   1.4 &   0.12 &   0.25 &   0.14 &   1.6
\\ 
{\bf d}  &  12 &   1.5 &   0.92 &   0.06 &   0.14 &   0.04 &   4.7
\\ 
\hline
\end{tabular}
\end{center}
\caption{Cross-sections (in fb) for signal and main background processes with the generator-level cuts as 
in Eq.~(\ref{eq:gen}) (first row) and the
event selection cuts presented in Section~\ref{sec:kinematics}. 
The last column displays the resulting signal-to-background ratios. 
\label{tab:evsel}}
\end{table}

\begin{figure}[t]
\vspace*{0.2cm}
\centerline{\includegraphics[height=7.8cm,angle=0,clip=true]{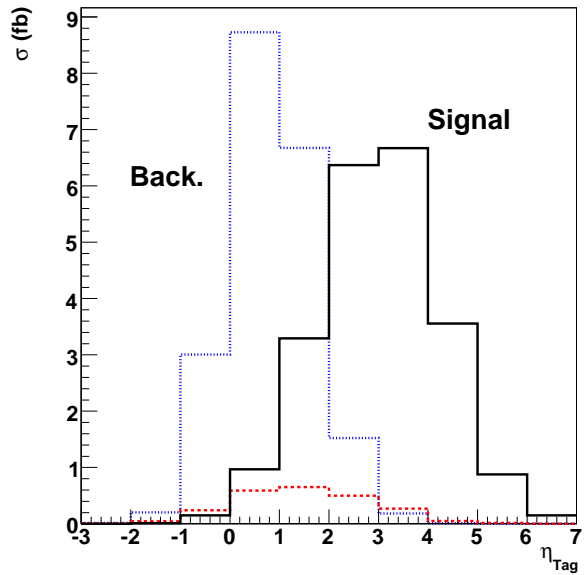}}
\caption{Pseudorapidity distribution of the forward tag parton after the application of Cuts {\bf a-b} (see Section~\ref{sec:kinematics}). The solid, dotted and dashed histograms correspond to the Higgs boson signal, $e^-p\rightarrow \nu_e t\overline{b}+X$, $e^-p\rightarrow \nu_e b\overline{b}j+X$ processes, respectively. Results are given in units of fb per bin.
  \label{fig:etaj}}
\end{figure}

Figure~\ref{fig:etaj} displays the pseudorapidity distributions for the forward tag parton after the application of Cuts {\bf a-b}. The solid histogram corresponds to the Higgs boson signal. The kinematics of the forward tag parton in signal resemble that of the Higgs boson 
simulated for VBF in $pp$ collisions. The forward tag parton points predominantly along the direction of the incoming proton.
The dotted and dashed histograms correspond to the forward tag parton for the two leading background processes. The forward tag parton due to the process $e^-p\rightarrow \nu_e t\overline{b}+X$ arises from the decay of the $W$ boson, which is predominantly produced centrally. 
The kinematics arising from CC production of $Z$ bosons differs qualitatively due to the presence of transversely 
polarized $W$'s. In this case the momentum transfer at the $fWf^{\prime}$ vertex is significantly larger, and therefore, the scattered quark is more central than in the case of the Higgs boson. For the same reason the observed $\met$ is significantly larger in this process.

\begin{figure}[t]
\vspace*{0.2cm}
\centerline{\includegraphics[height=7.8cm,angle=0,clip=true]{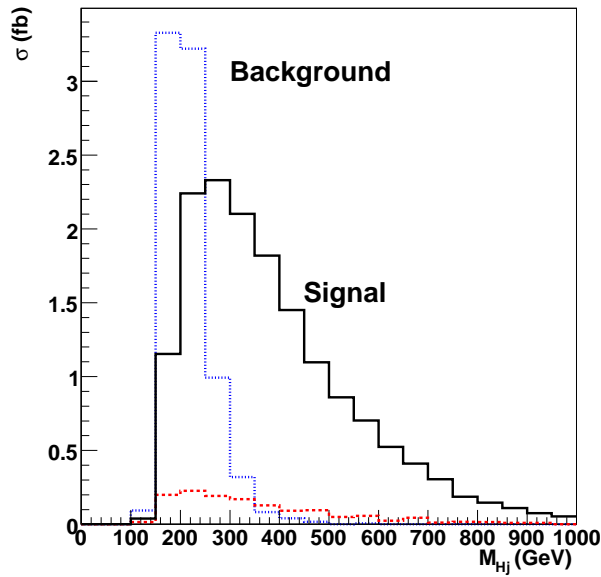}}
\caption{Distribution of the invariant mass of the Higgs boson candidate and the forward tag parton after the application of Cuts {\bf a-c} (see Section~\ref{sec:kinematics}). The solid, dotted and dashed histograms correspond to the Higgs boson signal, $e^-p\rightarrow \nu_e t\overline{b}+X$, $e^-p\rightarrow \nu_e b\overline{b}j+X$ processes, respectively. 
Results are given in units of fb per bin.
  \label{fig:mhj}}
\end{figure}

Figure~\ref{fig:mhj} shows the distribution of the invariant mass of the Higgs boson candidate and the forward tag parton. The solid, dotted and dashed histograms correspond to the Higgs boson signal, $e^-p\rightarrow \nu_e t\overline{b}+X$, $e^-p\rightarrow \nu_e b\overline{b}j+X$ processes, respectively. The same discussion mentioned for Figure~\ref{fig:etaj} applies here. The variable shown in Fig.~\ref{fig:mhj} is particularly effective in suppressing the $e^-p\rightarrow \nu_e t\overline{b}+X$ process.\footnote{In this process the pseudorapidity difference between the Higgs boson candidate and the forward tag parton is significantly smaller than that of the signal. However, the invariant mass of the forward tag parton and the Higgs boson candidate remains a better discriminator.}

\begin{figure}[t]
\vspace*{0.2cm}
\centerline{\includegraphics[height=7.8cm,angle=0,clip=true]{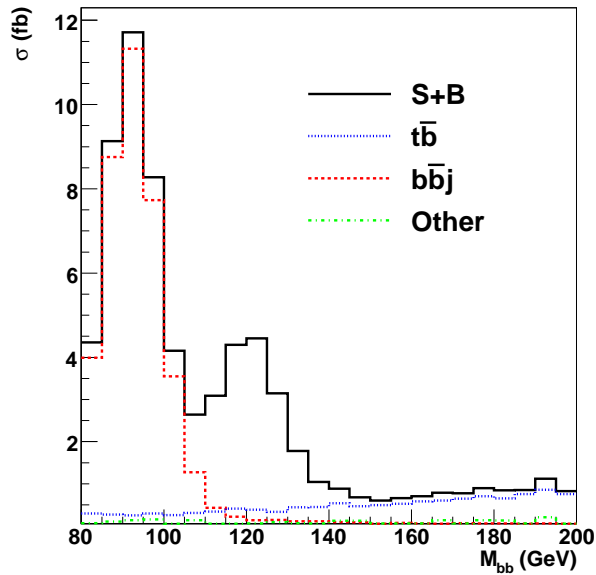}}
\caption{Invariant mass spectrum of the Higgs boson candidate after the application of Cuts {\bf a} and {\bf c-d} (see Section~\ref{sec:kinematics}). The black histogram corresponds to the sum of the signal and background processes. The contribution from the different background processes is given in separate histograms (see text). Results are given in units of fb per bin.
  \label{fig:mbb}}
\end{figure}

Figure \ref{fig:mbb} presents the invariant mass spectrum of the Higgs boson candidates after the application of Cuts {\bf a} and {\bf c-d} (excluding Cut {\bf b}). The solid histogram in Fig.~\ref{fig:mbb} corresponds to the sum of the signal and background processes. The contribution from the two leading backgrounds, $e^-p\rightarrow \nu_e W^-b\overline{b}+X$, $e^-p\rightarrow \nu_e b\overline{b}j+X$, are given by the dark dotted and dashed histograms, respectively. The total contribution from the rest of the processes is given by the light dotted histogram.

Table~\ref{tab:evsel} displays the predicted cross-sections (in fb) 
for signal and the main backgrounds after the application of the various cuts discussed above. 
The last column of Table~\ref{tab:evsel} illustrates the significant enhancement of the signal-to-background ratio after the application of Cuts {\bf c-d}. 
It is important to note that the background efficiencies for some of the backgrounds are reported here after requiring the presence of an additional parton with a $p_T$ cut. Therefore the signal-to-background ratios reported after Cuts {\bf a-b} are not representative of those for a purely inclusive analysis. For instance, the cross-section of the photo-production process $e^-p\rightarrow e^-b\overline{b}+X$ after application of Cuts {\bf a-b} is about 3.5 times larger than that of  the photo-production process $e^-p\rightarrow e^-b\overline{b}j+X$ with the generator cuts specified in Section~\ref{sec:background}.

The impact of the central jet veto on the signal production is also considered here. We evaluate the additional suppression power of imposing a veto on events with additional partons in top-quark related backgrounds. In the processes $e^-p\rightarrow W^-b\overline{b}+X$ and $e^-p\rightarrow e^-t\overline{t}+X$ the forward tag parton comes from the decay of the $W$ boson, which implies the presence of another parton in the event. By imposing a veto on events with an additional parton with $p_T>30\,\gev$ in the range $\left|\eta\right|<5$ the two backgrounds drop by 40\,\% and 50\,\%, respectively. The loss of signal  is about  $7 \%$. 

It is important to evaluate the dependence of the Higgs boson signal on the energy of the incoming electron. Table~\ref{tab:energy} gives a summary of the Higgs boson signal efficiencies for different values of $E_e$.  The first rows correspond to the cumulative efficiencies and the second rows show the efficiencies for that cut with respect to the previous cut. The transverse momenta of the Higgs boson, its decay products and the forward tag parton do not depend strongly on $E_e$. The lower the energy is, the larger the longitudinal boost acquired by the Higgs boson, and, therefore, the decay products become more forward in the laboratory frame. This explains most of the loss of signal efficiency after Cut {\bf a} as $E_e$ decreases. 
%
As $E_e$ decreases, the gauge boson off the quark line needs to be more energetic to produce a Higgs boson
and the accompanying forward parton carries less energy. 
Also, the Higgs boson and the forward parton appear closer to each other in the laboratory frame and 
their invariant mass decreases at a lower $E_e$. This is illustrated by the $E_e$ dependence of the signal efficiencies after Cut {\bf d} with respect to that after Cut {\bf c}. A similar effect is expected in the relevant background processes. 
Overall, the efficiency of the signal has a tendency to decrease as $E_e$ decreases for $E_e<100\,\gev$, where for $E_e>100\,\gev$ the dependence on $E_e$ is not significant.

\begin{table}[t]
\begin{center}
\begin{tabular}{|c||c|c|c|c||}
\hline
Cut & $E_e=50$ & $E_e=100$  & $E_e=140$ &  $E_e=200$ \\ 
\hline
{\bf a}  &  0.129 &  0.157 &  0.166 &  0.171
\\ 
  & -  & -  &  -  & - 
\\ 
\hline 
{\bf b}  &  0.109 &  0.127 &  0.132 &  0.136
\\ 
  &   0.84 &   0.81 &   0.80 &   0.80
\\ 
\hline 
{\bf c}  &  0.076 &  0.090 &  0.093 &  0.095
\\ 
  &   0.70 &   0.71 &   0.70 &   0.70
\\ 
\hline 
{\bf d}  &  0.050 &  0.067 &  0.073 &  0.078
\\ 
  &   0.66 &   0.75 &   0.79 &   0.82
\\ 
\hline 
\end{tabular}
\end{center}
\caption{Higgs boson signal efficiencies for different energies of the incoming electron (in \gev) using the event selection chosen in Section~\ref{sec:kinematics}. The first rows correspond to the cumulative efficiencies. The second rows show the efficiencies with respect to the previous cut. \label{tab:energy}}
\end{table}

A check was performed with a degraded scenario for the hadronic energy resolution as in Eq.~(\ref{eq:res}), 
$\alpha = 0.7\gev^{1 \over 2}$ and $\beta=0.05$. The relative resolution on the reconstructed Higgs boson mass degrades to $9\%$. 
The signal efficiency in the mass window degrades by $17\%$ 
while the three leading backgrounds increase substantially. The contribution from the photo-production process $e^-p\rightarrow b\overline{b}j+X$ increases by about an order of magnitude and that of the process $e^-p\rightarrow \nu_e b\overline{b}j+X$ by almost a factor of two. Overall, the signal-to-background ratio degrades by about a factor of two. This indicates that the hadronic energy resolution is an essentidal element of the detector performance.

High performance of $b$-tagging is crucial as well to the robustness of the Higgs boson search. The fake backgrounds considered here come from the process $e^-p\rightarrow \nu_e jjj+X$ for which the final estate is composed of an admixture of light quarks and $c$-jets. With the fake rejection assumed here this background, which is dominated by $c$-quark fakes, constitutes less than $5\%$ of the total background. The analysis could be optimized with a looser $b$-tagging efficiency without incurring in a significant increase of the total background.

The impact of the extension of the acceptance of the tracking system is evaluated. The increase of the tracking acceptance to $\left|\eta\right|<3$ enhances the signal yield by about $6\%$ while reducing the signal-to-background ratio by about 
$4\%$. A further increase of the tracking acceptance to $\left|\eta\right|<3.5$ enhances the signal yield by only $1.5\%$. 

\section{Neutral Current Signal}
\label{sec:nc}

The NC process has the advantage that the electron reconstruction is superior with respect to that of the missing neutrino in the CC process. 
However, it is not background-free.
Furthermore, the NC process has a significantly smaller cross-section with respect to the CC process 
as seen in  Fig.~\ref{fig:tot}.

The leading background arises from $e^-p\rightarrow e^-b\overline{b}j+X$, largely from the reactions
$\gamma/Z\ g \to b\bar b$.  
To evaluate the signal-to-background ratio,  we exploit the following acceptance cuts. We demand the existence of a
backward electron and a pair of $b$ quarks
\begin{equation}
p_T(e) > 30\ \gev,\quad |\eta_e| < 5,\quad  p_T(b) > 30\ \gev,\quad |\eta_b| < 2.5.
\end{equation}
The invariant mass of the $b$-quarks are required to be in the same mass window as that considered in Section~\ref{sec:cc}. 
We also require a forward parton in
\begin{equation}
p_T(j) > 30\ \gev,\quad |\eta_j| < 5.
\end{equation}
%
Assuming the same $b$-jet tagging efficiciency as used in Section~\ref{sec:cc} and selecting the $b\bar b$ events
within the Higgs mass window as before, the signal and background cross-sections are 5.7\,fb and 23.7\,fb, 
respectively, leading to a signal-to-background ratio of about 0.25. This may yield a $4\sigma$ statistical significance
for the signal with an integrated luminosity of 10 fb$^{-1}$. 
The application of a cut on $M_{HJ}$ will improve the signal-to-background ratio, but at some cost of signal rate. 

We consider that the Higgs boson search using the NC production mechanism is an interesting prospect, as it has the potential to enhance the overall Higgs boson signal efficiency. More studies would be required to evaluate the sensitivity due to this final state. 

\section{Conclusions}
\label{sec:conclusions}

At the dawn of the LHC era, it is well motivated to consider the physics potential for the proposed 
proton-electron collider, the LHeC. 
We studied the use of forward jet tagging as a means to secure the observation of the 
Higgs boson in the $H\rightarrow b\overline{b}$ decay mode, and to significantly improve the purity of the signal.
An excellent signal-to-background ratio of almost a factor of five can be achieved for the CC process 
while allowing for a significant rate of Higgs boson events. 
With this we believe that a  measurement of  bottom Yukawa coupling at the LHeC may be feasible by means of combining the knowledge from the LHC on $H\rightarrow WW^{\star},\tau\tau$.

The implications of a veto on additional partons were explored. 
It was demonstrated that the $t$-quark related backgrounds can be further reduced 
by about a factor of two with a signal loss of about 7\%.

The dependence of the signal kinematics on the energy of the incoming electron were evaluated. 
Overall, the efficiency of the signal has a tendency to decrease as $E_e$ decreases for $E_e<100\,\gev$, where for $E_e>100\,\gev$ the dependence on $E_e$ is not significant.

Two detector performance aspects have been identified as essential to the Higgs boson search: hadronic energy resolution 
and $b$-tagging capabilities. When considering a more conservative scenario for the hadronic energy resolution the signal-to-background ratio in the Higgs boson search degrades by a factor of two. Large signal-to-background ratios can only be achieved with adequate $b$-tagging capabilities.

We also considered the isolation of the Higgs boson using the NC production mechanism. Although with a much
smaller signal cross section and substantial backgrounds, 
we demonstrated that this has the potential to enhance the overall Higgs boson signal observation.

Due to the small di-Higgs boson cross-sections, setting meaningful limits on the trilinear Higgs boson self-coupling 
is extremely challenging in the range of energies of $ep$ collisions considered here.

A simple cut-based analysis was performed here.  
More complex discriminators could be constructed in order to enhance the efficiency of the Higgs boson signal. 

\begin{acknowledgments}
We would like to thank J.~Alwall and F.~Maltoni for their invaluable help with the Madgraph package, A. de Roeck,  
T.~Vickey, and D.~Zeppenfeld for discussions. 
%
This work was supported in part by the DOE Grants No. DE-FG0295-ER40896 and DE-FG02-95ER40896. The work of B.M. is also supported by the Wisconsin Alumni Research Foundation.
\end{acknowledgments}

\bibliography{vbf,mycites,ephjprd}
\end{document}